\documentclass[times, 10pt]{article}
\usepackage{times}
\usepackage{amsfonts,latexsym, amssymb, amsmath}
\usepackage{proof, bussproofs}

\def\bc{\mathcal{B}}                  
\def\mc{\mathcal{M}}                  

\newcommand{\QQ}{\mathbb{Q}}

\def\lis#1#2{\mathsf{doors}(#1,#2)}

\def\prm#1{\mathbf{#1}}
\newcommand{\pref}{\leq}
\newcommand{\EALS}{$EAL^{\star}$}
\newcommand{\DLAL}{$DLAL$}
\newcommand{\DLALS}{ $DLAL{\star}$ }

\def\bparam#1{par^{bool}(#1)}
\def\inparam#1{par^{int}(#1)}
\def\lift#1{\overline{#1}}

\newcounter{pb}
\newtheorem{problem}[pb]{Problem}

\def\llto{\mathbin{-\mkern-3mu\circ}}              

\def\lltensor{\otimes}

\newcommand{\QED}{\hspace*{\fill}\vrule height6pt width6pt depth0pt}

\def\unif#1#2{\mathcal{U}(#1, #2)}        

\def\bracket{\mathsf{bracket}}
\def\wbracket{\mathsf{wbracket}}
\def\Bra{\mathsf{Bracket}}
\def\Bang{\mathsf{Bang}}
\def\Scope{\mathsf{Scope}}

\def\Ltype{\mathsf{Ltype}}
\def\Const{\mathsf{Const}}

\def\fCenter{{~ \vdash ~}}
\newcommand {\fli} {\Rightarrow}
\newcommand{\lf}{\ma{L}_{F}}              
\newcommand{\ldlal}{\ma{L}_{DLAL}}        
\newcommand{\ldlals}{\ma{L}_{DLAL\star}}  

\def\FV#1{FV({#1})}           

\newcommand{\ma}{\mathcal}

\def\eras#1{{#1}^-}                   

\newcommand{\ZZ} {\mathbb{Z}}

\def\sub#1#2 {[ #1 \slash #2 ]}                                            

\newcommand {\fm} {\multimap}

 \newcommand{\bs} {\mathord{!}}

  \newcommand{\pa} {\mathord{\S}}
\newcommand{\pad} {\mathord{\bar{\S}}} 

\newcommand{\al} {\alpha}
\newcommand{\la} {\lambda}
\newcommand{\La} {\Lambda}

\newtheorem{theo}{Theorem}

\newtheorem{prop}[theo]{Proposition}
\newtheorem{lem}[theo]{Lemma}

\title{Verification of Ptime reducibility for\\ system F terms via
 Dual Light Affine Logic}

\begin{document}
\author{Vincent Atassi\thanks{Work
         partially supported by projects CRISS (ACI), GEOCAL (ACI), NO-CoST (ANR)}\\
LIPN, Univ. Paris-Nord, France\\
atassi@lipn.univ-paris13.fr
\and
Patrick Baillot$^*$ \\
LIPN, Univ. Paris-Nord, France\\
pb@lipn.univ-paris13.fr
\and
Kazushige Terui \\
NII,
 Tokyo, Japan\\
terui@nii.ac.jp
}

\maketitle
\thispagestyle{empty}

\begin{abstract}

In a previous work we introduced Dual Light Affine Logic (DLAL)
(\cite{BaillotTerui04}) as a variant of Light Linear Logic suitable for
guaranteeing complexity properties on lambda-calculus terms: all typable
terms can be evaluated in polynomial time and all Ptime functions can
be represented. In the present work we address the problem of typing
lambda-terms in second-order DLAL. For that we give a procedure which,
starting with a term typed in system F, finds all possible ways to
decorate it into a DLAL typed term. We show that our procedure can be
run in time polynomial in the size of the original Church typed system
F term.
%
 \end{abstract}

\section{Introduction}


  Several works have
studied programming languages with intrinsic computational complexity
properties. This line of research, Implicit computational complexity
(ICC), is motivated both by the perspective of automated complexity
analysis, and by foundational goals, in particular to give natural
characterizations of complexity classes, like Ptime or
Pspace. Different calculi have been used for this purpose coming from
primitive recursion, lambda-calculus, rewriting systems  (\textit{e.g.}
\cite{BellantoniCook,MarionMoyen00,LeivantMarion93})\dots A
convenient way to see these systems is in general to describe them as
a subset of programs of a larger language satisfying certain criteria:
for instance primitive recursive programs satisfying safe/ramified
recursion conditions, rewriting systems admitting a termination
ordering and quasi interpretation, etc\dots

\textbf{Inference.} To use such ICC systems for programming purpose it is
natural to wish to automatize the verification of the criteria. This way
the user could stick to a simple programming language and the compiler
would check whether the program satisfies the criteria, in which case
a complexity property would be guaranteed.

 In
general this decision procedure involves finding a certain \textit{witness}, like a type,
a proof or a termination ordering.  Depending on the system this
witness might be useful to provide more precise information, like an
actual bound on the running time, or a suitable strategy to evaluate
the program. It might be used as a certificate guaranteeing a particular
quantitative property of the program.

\textbf{Light linear logic.} In the present work we consider the
 approach of Light linear logic (LLL) (\cite{Girard98}), a variant of
 Linear logic which characterizes polynomial time computation, within
 the proofs-as-programs correspondence. It includes higher-order and
 polymorphism, and can be extended to a naive set theory
 (\cite{Terui04}), in which the provably total functions correspond to
 the class of polynomial time functions.

The original formulation of LLL by Girard was quite complicated, but a
first simplification was given by Asperti with Light Affine Logic
(LAL)  (\cite{AspertiRoversi02}). Both systems have two modalities (one more than Linear logic)
to control duplication. There is a forgetful map to system F terms
(polymorphic types) obtained by erasing some information (modalities)
in types; if an LAL typed term $t$ is mapped to an F-typed term $M$ we
also say that $t$ is a \textit{decoration} of $M$.

 So an LAL program can be understood as a system F program, together
with a typing guarantee that it can be evaluated in polynomial
time. As system F is a reference system for the study of
polymorphically typed functional languages and has been extensively
studied, this seems to offer a solid basis to LAL.

 However LAL itself is still difficult to handle and following the
 previous idea for the application of ICC methods, we would prefer to
 use plain lambda-calculus as a front-end language, without having to
 worry about the handling of modalities, and instead to delegate the
 LAL typing part to a type inference engine. The study of this
 approach was started in \cite{Baillot02}. For it to be fully
 manageable however several conditions should be fulfilled:
\begin{enumerate}
\item  a suitable way to execute the lambda-terms with the expected
complexity bound,
\item  an efficient type inference,
\item a typed language which is expressive enough so that a reasonable 
range of programs is accepted.
\end{enumerate}

 The language LAL presents some drawback for the first point, because the LAL typed
terms need to be evaluated with a specific graph syntax, \textit{proof-nets}, in order
to satisfy the polynomial bound, and plain beta reduction can lead to
exponential blow-up. 
In a previous work (\cite{BaillotTerui04}) we
addressed this issue  by defining a subsystem of LAL, called
Dual Light Affine Logic (DLAL). It is defined with both linear and
non-linear function types. It is complete for Ptime just as LAL and
its main advantage is that it is also Ptime sound w.r.t. beta
reduction: a DLAL term admits a bound on the length of all its beta
reduction sequences. Hence DLAL stands as a reasonable substitute for
plain LAL for typing issues.

Concerning point 2, as type inference for system F is undecidable we
don't try to give a full-fledged type inference algorithm from untyped
terms.  Instead, to separate the polymorphic part issue from the
proper DLAL part one, we assume the initial program is already typed
in F. Either the system F typing work is left to the user, or one could use a
partial algorithm for system F typing for this preliminary phase.
 
 So the contribution of the present work is to define an efficient
algorithm to decide if a system F term can be decorated in a DLAL typed
term. This was actually one of the original motivations for defining
DLAL. We show here that decoration can be performed in polynomial
time. This is obtained by taking advantage of intuitions coming from
proof-nets, but it is presented in a standard form with a first phase
consisting in generating constraints expressing typability and a
second phase for constraints solving. 
One difficulty is that the
initial presentation of the constraints involves disjunctions of
linear constraints, for which there is no obvious Ptime bound. Hence
we provide a specific resolution strategy.

The complete algorithm is 
already implemented in ML, 
in a way that follows closely the specification given in the article.
It is modular and  
usable with any linear constraints solver.
The code is commented, and
available for public download (Section \ref{l-implement}).
 With this program one might thus write terms in system F and 
 verify if they are Ptime and obtain a time upper
bound.  It should in particular be useful to study further properties
of DLAL and to experiment with reasonable size programs.

 The point 3 stressed previously about expressivity of the system
remains an issue which should be explored further. Indeed the DLAL
typing discipline will in particular rule out some nested iterations
which might in fact be harmless for Ptime complexity. This is related
to the line of work on the study of intensional aspects of Implicit
computational complexity (\cite{MarionMoyen00,Hofmann03}). 

 However it might be possible to consider some combination of DLAL
with other systems which could allow for more flexibility, and we
think a better understanding of DLAL, and in particular of its type
inference, is a necessary step in that direction.

\textbf{Related work.} Inference problems have been studied for
several ICC systems (\cite{Amadio05,HofmannJost02}). Elementary
linear logic (EAL) in particular is another variant of Linear logic
which characterizes Kalmar elementary time and has applications to
optimal reduction. Type inference in the propositional fragment
of this system has been studied in \cite{CoppolaMartini01,CoppolaRonchi03,CoppolaDalLagoRonchi05}
and \cite{BaillotTerui05} which gives a polynomial time procedure. Type
inference for LAL was also investigated, in \cite{Baillot02,Baillot04}.
To our knowledge the present algorithm is however the first one for dealing
with polymorphic types in a EAL-related system, and also the first
one to infer light types in polynomial time.

\medskip

\textbf{Notations}. Given a lambda-term $t$,  $\FV{t}$ will be the
set of its free variables. 
 The prefix relation
on words will be denoted by $\pref$.


\section{From system F to \DLAL}

 The language $\lf$ of system F types is given by:
$$T, U::= \alpha \; | \; T \rightarrow U \;|\; \forall \al . T$$

We assume that a countable set of term variables
$x^T, y^T, z^T,\ldots$ is given for each type $T$.
The terms of system $F$ are built as follows
(here we write $M^T$ to indicate that the term $M$ has type $T$):
$$
x^T
\quad
(\la x^T. M^U)^{T\rightarrow U}
\quad
((M^{T\rightarrow U}) N^T)^U$$
$$
(\La \alpha. M^U)^{\forall \alpha.U}
\quad
((M^{\forall\alpha. U})T)^{U[T/\alpha]}
$$
with the proviso that when building a term
$\La \alpha. M^U$,
$\alpha$ may not occur freely in the types of
free term variables of $M$ (the {\em eigenvariable condition}).


It is well known that
there is no sensible resource bound (i.e. time/space) on the execution
of system F terms in general. To impose some bounds, a more refined
type system is required. \DLAL\ serves well as such a
type system.

 The language $\ldlal$ of DLAL types is given by:
$$A, B::= \alpha \; | \; A \fm B \; | \; A \fli B \; |\; \pa A   \;|\; \forall \al . A$$
We note $\pa^0 A=A$ and   $\pa^{k+1} A=\pa \pa^k A$.
 The erasure map $\eras{(.)}$ from $\ldlal$ to $\lf$ is defined by:
$$\eras{(\pa A)}= \eras{A},\ \ \
\eras{(A\fm B)}=\eras{(A\fli B)}=\eras{A} \rightarrow \eras{B},$$
and $\eras{(.)}$ commutes
to the other connectives.
 We say
$A\in \ldlal$ is a {\em decoration} of 
$T\in \lf$ if $\eras{A}=T$.

A {\em declaration} is a pair of the form $x^T: B$ with $B^- = T$. It
is often written as $x: B$ for simplicity.  A {\em judgement} is of
the form $\Gamma;\Delta \vdash M: A$, where $M$ is a system F term,
$A\in \ldlal$ and $\Gamma$ and $\Delta$ are disjoint sets of
declarations.  When $\Delta$ consists of $x_1:A_1, \ldots, x_n:A_n$,
$\pa\Delta$ denotes $x_1: \pa A_1, \ldots, x_n:\pa A_n$.  The type
assignment rules are given on Figure \ref{NDLALrules}.  Here, we
assume that the substitution $M[N/x]$ used in ($\pa$ e) is {\em
capture-free}. Namely, no free type variable $\alpha$ occurring in $N$
is bound in $M[N/x]$.  We write $\Gamma;\Delta \vdash_{DLAL} M: A$ if
the judgement $\Gamma;\Delta \vdash M: A$ is derivable.

 \begin{figure*}[ht]
\begin{center}
\fbox{
\begin{tabular}{c@{}cc}
  &{\infer[\mbox{(Id)}]{;x^{A^-}:A \vdash x^{A^-}:A}{}} & \\
 &&\\
 &{\infer[\mbox{($\fm$ i)}]{\Gamma;\Delta \vdash \la x^{A^-}.
M: A \fm B }
 {\Gamma;x:A,\Delta \vdash M:B}}
  & {\infer[\mbox{($\fm$ e)}]{\Gamma_1,\Gamma_2;
\Delta_1,\Delta_2\vdash (M) N :B }
  {\Gamma_1;\Delta_1 \vdash M:A \fm B & \Gamma_2;\Delta_2 \vdash N:A}}\\[1ex]
&{\infer[\mbox{($\fli$ i)}]{\Gamma;\Delta \vdash \la x^{A^-}. M: A \fli B }
 {x: A, \Gamma;\Delta  \vdash M:B}}
  & {\infer[\mbox{($\fli$ e) (*)}]{\Gamma, z: C;\Delta \vdash (M) N :B }
  {\Gamma;\Delta \vdash M:A \fli B &  ;z:C \vdash N:A}}\\[1ex]
&{\infer[\mbox{(Weak)}]{\Gamma_1,\Gamma_2;\Delta_1,\Delta_2 \vdash M: A }
 {\Gamma_1;\Delta_1 \vdash M:A & }}
  &{\infer[\mbox{(Cntr)}]{x: A, \Gamma;\Delta \vdash M[x \slash x_1, x \slash x_2] :B }{x_1: A,x_2: A, \Gamma;\Delta \vdash M:B }} \\[1ex]
&{\infer[\mbox{($\pa$ i)}]{\Gamma; \pa\Delta \vdash M: \pa A }
 { ;\Gamma, \Delta \vdash M:A}}
  & {\infer[\mbox{($\pa$ e)}]{\Gamma_1,\Gamma_2;\Delta_1,\Delta_2
     \vdash M[N \slash x] :B }
  {\Gamma_1;\Delta_1 \vdash N: \pa A  & \Gamma_2;x:\pa A,\Delta_2
\vdash M:B}}\\[1ex]
&{\infer[\mbox{($\forall$ i) (**)}]{\Gamma;\Delta \vdash  \La \alpha.M:\forall \alpha. A}{\Gamma;\Delta \vdash M:A}} &
 {\infer[\mbox{($\forall$ e)}]{\Gamma;\Delta \vdash (M)\eras{B} :A[B \slash \al] }
{\Gamma;\Delta \vdash M:\forall \al. A}} \\[1ex]
&\multicolumn{2}{c}{
\mbox{(*) $z:C$ can be absent.}}\\
&\multicolumn{2}{c}{
\mbox{(**) $\alpha$ does not occur freely in $\Gamma$.}}
 \end{tabular}
}
  \caption{Typing system F terms in \DLAL}\label{NDLALrules}
\end{center}
\end{figure*}

Recall that binary words, in  $\{0,1\}^*$,  can be given the following type
in F:
 $$W_F
 = \forall\alpha.
(\alpha\rightarrow\alpha)\rightarrow
(\alpha\rightarrow\alpha)\rightarrow
 (\alpha\rightarrow\alpha)$$
A corresponding type in DLAL, containing the same terms, is given by:
$$W_{DLAL} = \forall\alpha.
(\alpha\llto\alpha)\Rightarrow
(\alpha\llto\alpha)\Rightarrow
\pa (\alpha\llto\alpha)$$

The {\em depth} $d(A)$ of a \DLAL\ type $A$ is defined by:
\[
\begin{array}{rcl}
d(\alpha) & = & 0, \qquad \qquad \qquad \qquad d(\forall\alpha. B)  =  d(B), \\
d(A\fm B) & = & max(d(A),d(B)), \quad d(\pa A)  =  d(A)+1,\\
d(A\fli B) &=&  max(d(A)+1,d(B)).
\end{array}
\]
A type $A$ is said to be $\Pi_1$ if it does not
contain a negative occurrence of $\forall$; like for instance
$W_{DLAL}$.

The fundamental properties of \DLAL\ are the following
\cite{BaillotTerui04}:
\newpage
\begin{theo}~
\begin{enumerate}
\item
Let $M$ be a closed term of system F that has a $\Pi_1$ type $A$ in
\DLAL. Then
$M$ can be normalized in $O(|M|^{2^d})$ steps by $\beta$-reduction,
where $d=d(A)$ and $|M|$ is the structural size of $M$.
\item Every Ptime function $f:\{0,1\}^* \longrightarrow \{0,1\}^*$
can be represented by a closed term $M$ of type
$W_{DLAL} \llto \S^d W_{DLAL}$ for some $d\geq 0$.
\end{enumerate}
\end{theo}
Notice that the result 1 holds neither for Light linear logic
nor  Light affine logic.
Although they are logics of
polynomial time, they require some special proof syntax
such as proof nets \cite{Girard98,AspertiRoversi02}
or light affine lambda calculus
\cite{Terui01}
to guarantee polynomial time bounds.

The result 1 implies that if we ignore the embedded types occurring in
 $M$, the normal form of $M$ can be computed in polynomial time, when
 the depth is fixed.  It moreover shows that a term $M$ of type $W_{DLAL}\fm
 \pa^d W_{DLAL}$ is Ptime, because then for any Church word
 $\underline{w}$ we have that $(M)\; \underline{w}$ has type
 $\pa^d W_{DLAL}$, and can thus be evaluated in time
 $O(|\underline{w}|^{2^{d+1}})$.
  
The result 2 on the other hand guarantees that \DLAL\ has sufficient expressive power,
at least enough to (extensionally) represent all polynomial time
functions.


Now, let $M^{W_F\rightarrow W_F}$ be a system F typed term and suppose
that we know that it has a \DLAL\ type $W_{DLAL} \llto \pa^d W_{DLAL}$
for some $d\geq 0$.  Then, by the consequence of the above theorem, we
know that the term $M$ is Ptime.
Thus by assigning \DLAL\ types to a given system F term, one can
 {\em statically verify} a polynomial time bound for
its execution.

As a first step to elaborate this idea to use \DLAL\ for resource
verification of system F terms, we address the following:
\begin{problem}[\DLAL\ typing]
Given a closed term $M^T$ of system F, determine if there
is a decoration $A$ of $M$
such that
$\vdash_{DLAL} M:A$.
\end{problem}
(Here the closedness assumption is only for readability.)

In the sequel, we show that there is
a polynomial time algorithm for solving
the \DLAL\ typing problem. 

 This should be contrasted with the fact that
the set of system F terms representing Ptime functions is not 
recursively enumerable (this can be easily proved by reduction of
Hilbert's 10th problem).

Hence even though \DLAL\ does not capture all Ptime terms, the general
problem is undecidable and this type system gives a partial but
efficiently realizable verification method.

\section{Characterizing \DLAL\ typability}

\subsection{Pseudo-terms}

To address the \DLAL\ typing problem, it is convenient to introduce
an intermediary syntax which is more informative than system F terms
(but not more informative than \DLAL\ derivations themselves).

First we decompose $A\fli B$ into $\bs A\fm B$.
The language $\ldlals$ of \DLALS\ types is given by:
\begin{eqnarray*}
A & ::= & \alpha\ |\ D\fm A\ |\ \forall\alpha.A \ |\ \pa A \\
D & ::= & A\ |\ \bs A
\end{eqnarray*}
There is a natural map $(.)^\star$
from $\ldlal$ to $\ldlals$
such that $(A\fli B)^\star = \bs A^\star \fm B^\star$
and commutes with the other operations. 
The erasure map $\eras{(.)}$ from $\ldlals$ to $\lf$ can be
 defined as before.
A \DLALS\ type  is called a {\em bang type} if
it is of the form $\bs A$,
and otherwise called a {\em linear type}.
In the sequel, $A,B,C$ stand for linear types, and $D,E$ for
either bang or linear types.

We assume there is a countable set of term variables
$x^D, y^D, z^D,\dots$ for each $D\in$ \DLALS.
The \textit{pseudo-terms} are defined by the following grammar:
$$ t, u::= x^{D} \;|\; \la x^{D}. t \;|\;
(t)u \;|\;
 \La \alpha.t \;|\; (t)A \;|\; \pa t \;|\; \pad t,$$
where $A$ is a linear type and $D$ is an arbitrary one.
 The idea is that
$\pa$
corresponds to the main door of a $\pa$-box (or a $\bs$-box)
in \textit{proof-nets} (\cite{Girard87a, AspertiRoversi02}) while
$\pad$
 corresponds to auxiliary doors. But note
that there is no information in the pseudo-terms to link occurrences
of $\pa$ and $\pad$ corresponding to the same box, nor distinction
between $\pa$-boxes and $\bs$-boxes.

 There is a natural erasure map from pseudo-terms to
system F terms, which we will also denote
by $(.)^-$, consisting in removing all occurrences of $\pa$,
replacing $x^{D}$ with $x^{D^-}$ and $(t)A$ with $(t)A^-$.
When $t^- = M$, $t$ is called a
\textit{decoration} of $M$.

For our purpose, it is sufficient to consider
the class of \textit{regular} pseudo-terms, given by:
$$
t ::= x^{D}  \;|\;
 \la x^{D}. t \;|\;
(t)u \;|\;
 \La \alpha.t \;|\; (t)A \;|\;  \\
  \pa^m t,
$$
where $m$ is an arbitrary value in $\ZZ$ and $\pa^m t$ denotes
$\underbrace{\pa\cdots \pa}_{m\ times} t$ if
if $m\geq 0$, and 
$\underbrace{\pad\cdots \pad}_{-m\ times} t$ otherwise.
In other words, a pseudo-term is regular if
and only if it does not contain any subterm of the
form $\pa \pad u$ or $\pad \pa u$.

\subsection{Local typing condition}

We now try to assign types to pseudo-terms in a locally compatible way.
A delicate point
in \DLAL\ is that it is sometimes natural to associate {\em two}
types to one variable $x$. For instance, we have
$x:A; \vdash_{DLAL} x:\pa A$ in \DLAL, and this can be read as
$x:\bs A\vdash x:\pa A$ in terms of \DLALS\ types.
We thus distinguish between the 
{\em input types}, which are inherent to variables,
and the {\em output types}, which are
inductively assigned to
all pseudo-terms.
The condition (i) below is concerned with
the output types.
In the sequel,
$D^\circ$ denotes $\pa A$ if
$D$ is of the form $\bs A$,
and otherwise  denotes $D$ itself.

A pseudo-term $t$ satisfies the
{\em local typing condition}
if the following holds:
\begin{itemize}
\item[(i)] one can inductively assign a {\em linear} type
(called {\em the output type}) to each
subterm of $t$ in the following way (here the notation $t_A$
indicates that $t$ has the output type $A$):
$$
(x^{D})_{D^\circ}\quad\quad
(\pa t_A)_{\pa A}\quad\quad
(\pad t_{\pa A})_{A}\quad\quad
(\la x^{D}. t_B)_{D\fm B}
$$
$$
((t_{D\fm B})u_{D^\circ})_B\quad\quad
 (\La \alpha.t_A)_{\forall\alpha.A}\quad\quad
((t_{\forall\alpha.A})B)_{A[B/\alpha]}
$$
\item[(ii)] when a variable $x$ occurs more than once in $t$,
it is typed as $x^{\bs A}$,
\item[(iii)] $t$ satisfies the eigenvariable condition.
\end{itemize}
We also say that $t$ is {\em
locally typed}.

Notice that when $D$ is a bang type,
there is a type mismatch between $D$ and $D^\circ$
in the case of application. For instance, $(t_{\bs A\fm B})u_{\pa A}$
satisfies (i) whenever $t$ and $u$ do.
This mismatch will be settled by the bang condition
below. Observe also that the local typing rules are syntax-directed.

\subsection{Boxing conditions}

We now
recall definitions and results from \cite{BaillotTerui05}
giving some necessary conditions for a pseudo-term
to be typable (in \cite{BaillotTerui05} these conditions are used
for Elementary Affine Logic typing).
We consider words over the language $\mathcal{L}=\{\pa,\pad\}^{\star}$.
If $t$ is a pseudo-term and $u$ is an occurrence of subterm
 in $t$, let $\lis{t}{u}$ be the word inductively defined as follows:
$$\begin{array}{lcl}
\mbox{if $t=u$:}&& \lis{t}{u} = \epsilon, \\
\mbox{else: }&&\\
\lis{\pa t}{u} &=& \pa ::(\lis{t}{u}) \\
 \lis{\pad t}{u} &=& \pad::(\lis{t}{u})\\
 \lis{\la y^{D}.t_1}{u} &=&
\lis{\La \alpha.t_1}{u}\\
& =&
\lis{(t_1)A}{u} = \lis{t_1}{u} \\
\lis{(t_1)t_2}{u} &=& \lis{t_i}{u} \mbox{ where $t_i$ is the}\\
&&\mbox{ subterm
containing $u$}.
\end{array}
$$
That is to say, $\lis{t}{u}$ collects the modal symbols
$\pa$, $\pad$ occurring on the path from the root to
the node $u$
in the term tree of $t$.
 We define a map: $s: \mathcal{L} \rightarrow \ZZ$ by:
$$
s(\epsilon) = 0,\quad\quad
s(\pa :: l) = 1+ s(l),\quad\quad
s(\pad :: l) = -1+ s(l).
$$
A word $l\in\mathcal{L}$ is {\em weakly well-bracketed} if
$
\forall l'\leq l, s(l')\geq 0,
$
and
is {\em well-bracketed} if this condition holds and moreover $s(l)=0$.

\textbf{Bracketing condition.}
Let $t$ be a pseudo-term. We say that $t$ satisfies the
\textit{bracketing condition} if:
\begin{itemize}
\item[(i)] for any occurrence of free variable $x$ in $t$,
$\lis{t}{x}$ is well-bracketed;
\end{itemize}
moreover for any occurrence of
an abstraction subterm $\la x.v$ of $t$,
\begin{itemize}
\item[(ii)] $\lis{t}{\la x.v}$ is weakly well-bracketed, and
\item[(iii)] for any occurrence of $x$ in $v$,
$\lis{v}{x}$ is well-bracketed.
\end{itemize}

This condition is sufficient to rule out the canonical morphisms for
dereliction and digging, which are not valid in \DLAL\ (nor in $EAL$):
$$
(\la x^{\pa A}. \pad x)_{\pa A\fm A}\quad\quad
(\la x^{\pa A}. \pa x)_{\pa A\fm \pa\pa A}$$
Since $\lis{\pad x}{x}=\pad$ and $\lis{\pa x}{x}=\pa$, they do not
satisfy the bracketing condition (iii).

\textbf{Bang condition.} A subterm $u$ is called a {\em bang subterm}
of $t$ if it occurs as $(t'_{\bs A\fm B})u_{\pa A}$ in $t$.  
We say that a locally typed pseudo-term $t$ satisfies the
\textit{bang condition} if for any bang subterm $u$ of $t$,
\begin{itemize}
\item[(i)]  $u$ contains at most one free variable $x^{\bs C}$,
having a bang type $\bs C$.
\item[(ii)] for any  subterm $v$ of $u$ such that
$v\neq u$ and $v\neq x$,
$s(\lis{u}{v})\geq 1$.
\end{itemize}
This condition is sufficient to rule out the canonical morphisms for
monoidalness $!A\lltensor !B\llto !(A\lltensor B)$ and $\pa A\llto !A$
which are not valid in $LAL$
(the following terms and types are slightly more complicated since $\ldlals$
does not explicitly contain a type of the form $A\llto \bs B$):
$$
\lambda x^{\bs (A\fm B)}.\lambda y^{\bs B\fm C}. \lambda z^{\bs A}.
 (y)\pa((\pad x)\pad z)
$$
$$
\lambda x^{\pa A}.\lambda y^{\bs A\fm B}. (y)\pa (\pad x)
$$
In the first pseudo-term,
the bang subterm $\pa((\pad x)\pad z)$ contains more than one free variables.
In the second pseudo-term, the bang subterm
$\pa (\pad x)$ contains a free variable typed by a linear type.
Hence they both violate the bang condition (i).

\textbf{$\La$-Scope condition.}
The previous conditions, bracketing and bang, would be enough
to deal with boxes in the propositional fragment of $DLAL$.
For handling second-order quantification though, we need
a further condition to take into account the sequentiality enforced
by the quantifiers. For instance consider the following two formulas
(the second one is known as {\em Barcan's formula}):
\begin{eqnarray}
\pa \forall \alpha.  A \fm \forall \alpha. \pa A&&\\
\forall \alpha. \pa A \fm \pa \forall \alpha. A&&
\end{eqnarray}
 Assuming $\alpha$ occurs freely in $A$, formula (1) is provable while (2)
is not. Observe that we can build the following
 pseudo-terms which are locally typed and
have respectively type (1) and (2):
\begin{eqnarray*}
t_1&=& \la x^{\pa\forall\alpha.A}.\La \al. \pa((\pad x)\alpha)\\
t_2&=& \la x^{\forall\alpha.\pa A}.\pa \La \al. \pad((x)\alpha)
\end{eqnarray*}
 Both pseudo-terms satisfy the previous conditions, but
$t_2$ does not correspond to a $DLAL$ derivation.

Let $u$ be a locally typed pseudo-term.
We say that $u$ {\em depends on} $\alpha$ if the type of $u$ contains a free
variable $\alpha$.
We say that a locally typed pseudo-term $t$ satisfies the
\textit{$\La$-scope condition} if: for any subterm $\Lambda \al.u$ of
$t$ and for any subterm $v$ of $u$ that depends on $\al$,
$\lis{u}{v}$ is weakly well-bracketed.

Coming back to our example: $t_1$ satisfies the $\La$-scope condition,
but $t_2$ does not, because $(x)\alpha$ depends on $\alpha$ and nevertheless
$\lis{\pad((x)\alpha)}{(x)\alpha} = \pad$ is
not weakly well-bracketed.

\subsection{Correctness of the conditions}

\begin{prop}\label{p-correct}
If $M$ is a system F term such that the following judgement holds in DLAL:
$$(*)\quad x_1: A_1,\ldots, x_m : A_m;
y_1: B_1,\ldots, y_n : B_n \vdash M: C,$$
then there is a decoration $t$ of $M$ with type $C^\star$ and with
free variables $x_1^{\bs A_1^\star},
\ldots, x_m^{\bs A_m^\star}$,
$y_1^{B_1^\star},\ldots, y_n^{B_n^\star}$
which is
regular and satisfies the local
typing, bracketing, bang and $\La$-scope conditions.
\end{prop}
 
See the Appendix for the proof.

We want now to examine the converse property. 
First observe that whenever pseudo-terms
$\la x^{D}. t$,
$(t)u$,
$\La \alpha.t$, $(t)A$ satisfy
the
local typing, bracketing, bang and $\La$-scope conditions,
so do the immediate subterms $t$ and $u$.
The case of $\pa t$
is handled by the following key lemma
(already used for
\EALS\ in \cite{BaillotTerui05}):
\begin{lem}[Boxing]\label{boxinglemma}
If $\pa (t_A)$ is a pseudo-term which satisfies the local typing,
bracketing, bang and $\La$-scope
 conditions,
then there exist $v_A$, $(u_1)_{\pa B_1}$, \dots, $(u_n)_{\pa B_n}$,
unique (up to renaming
of $v$'s free variables) such that:
\begin{enumerate}
\item $FV(v)=\{x_1^{B_1}, \dots, x_{n}^{B_{n}}\}$ and
each $x_i$ occurs exactly once in $v$,
\item $\pa t=
\pa v[\pad u_1/x_1,\dots,\pad u_n/x_n]$
(substitution is assumed to be capture-free),
\item $v, u_1,\dots, u_n$ satisfy the same
conditions.
\end{enumerate}
\end{lem}

\begin{proof} Similar to the proof of Lemma 5
in \cite{BaillotTerui05}. See the Appendix.
\end{proof}

Thanks to the previous lemma, we can now prove:

\begin{theo}\label{t-correct}
Let $M$ be a system F term. Then
$x_1: A_1,\ldots, x_m : A_m;
y_1: B_1,\ldots, y_n : B_n \vdash M: C$ is derivable in
\DLAL\ if and only if
there is a decoration $t$ of $M$ with type
$C^\star$ and with
free variables $x_1^{\bs A_1^\star},
\ldots, x_m^{\bs A_m^\star}$,
$y_1^{B_1^\star},\ldots, y_n^{B_n^\star}$
which is
regular and satisfies the local
typing, bracketing, bang and $\La$-scope conditions.
\end{theo}

See Appendix \ref{s-proof} for the proof.
As a consequence, our \DLAL\ typing problem boils down to:

\begin{problem}[decoration]\label{decorationproblem}
Given a system F term $M$, determine if there exists
a decoration $t$ of $M$ which
is regular and satisfies the local typing, bracketing,
bang and $\Lambda$-scope conditions.
\end{problem}

\section{Parameterization and constraints}

\subsection{Parameterized terms and instantiations}

To solve the decoration problem (Problem \ref{decorationproblem}), 
one needs to explore the infinite set of decorations.
This can be effectively done 
by introducing an abstract kind of types and terms 
with symbolic parameters, and expressing the conditions for
such abstract terms to be materialized by boolean and
integer constraints over those parameters
(like
in the related type inference algorithms for EAL or LAL mentioned in the introduction).

We use two sorts of parameter: {\em integer parameters} $\prm{n}, \prm{m},\dots$
meant to range over $\ZZ$,
and {\em boolean parameters} $\prm{b_1}, \prm{b_2},\dots$ meant to range over
$\{0,1\}$. We also use
{\em linear combinations of integer parameters}
$\prm{c} = \prm{n_1} + \cdots + \prm{n_k}$, where $k\geq 0$ and each $\prm{n_i}$ is an
integer parameter. In case $k=0$, it is written as $\prm{0}$.

The set of {\em parameterized types} ({\em p-types} for short)
is defined by:
\begin{eqnarray*}
F & :: =&  \alpha\ |\
D\fm A \ |\
\forall\alpha. A \\
A & :: = & \pa^{\prm{c}} F \\
D & :: = & \pa^{\prm{b},\prm{c}} F
\end{eqnarray*}
 where
$\prm{b}$ is a boolean parameter and
$\prm{c}$ is a linear combination of integer parameters.
In the sequel, $A, B, C$ stand for {\em linear p-types}
of the form $\pa^{\prm{c}} F$, and $D$ for {\em bang p-types}
of the form $\pa^{\prm{b},\prm{c}} F$, and $E$ for arbitrary p-types.
When $D=\pa^{\prm{b},\prm{c}} F$, $D^\circ$
denotes the linear p-type $\pa^{\prm{c}} F$.
We assume that
there is a countable set of variables $x^{D},
y^{D}, \dots$
for each bang p-type $D$.
The \textit{parameterized pseudo-terms} ({\em p-terms} for short)
are defined by the following grammar:
\begin{eqnarray*}
t &::=& x^{D} \;|\; \la x^{D}. t \;|\;
(t) u \;|\;
 \La \alpha.t \;|\; (t)A \;|\; \pa^\prm{m} t.
\end{eqnarray*}

We denote by $\bparam{t}$
the set of boolean
parameters of $t$, and
by $\inparam{t}$ the set of integer parameters of $t$.

An {\em instantiation}
$\phi=(\phi^{b}, \phi^i)$ for a p-term $t$ is given
by two maps $\phi^b:\bparam{t}\rightarrow \{0,1\}$
 and $\phi^i:\inparam{t}\rightarrow \ZZ$.
The map $\phi^i$ can be naturally extended to linear combinations
$\prm{c} = \prm{n_1} + \cdots + \prm{n_k}$ by $\phi^i(\prm{c}) =
\phi^i(\prm{n_1}) + \cdots + \phi^i(\prm{n_k})$.  An instantiation
$\phi$ is said to be {\em admissible} for a p-type $E$ if for any
linear combination $\prm{c}$ occurring in $E$, we have
$\phi^i(\prm{c})\geq 0$, and moreover whenever $\pa^{\prm{b},\prm{c}}
F$ occurs in $E$, $\phi^b(\prm{b})=1$ implies $\phi^i(\prm{c})\geq 1$.
When $\phi$ is admissible for $E$, a type $\phi(E)$ of \DLALS\ is
obtained by replacing each $\pa^{\prm{c}} F$ and
$\pa^{\prm{b},\prm{c}} F$ with $\phi^b(\prm{b})=0$ by
$\pa^{\phi^i(\prm{c})} \phi(F)$, and $\pa^{\prm{b},\prm{c}} F$ with
$\phi^b(\prm{b})=1$ by $\bs\pa^{\phi^i(\prm{c})-1} \phi(F)$.

 So informally speaking, in $\pa^{\prm{b},\prm{c}} F$ the $\prm{c}$
 stands for the number of modalities ahead of the type, while the boolean
 $\prm{b}$ serves to determine whether the first modality, if any, is
 $\pa$ or $\bs$. 

An instantiation $\phi$ for a p-term $t$ is said to be {\em
admissible} for $t$ if it is admissible for all p-types occurring in
$t$.  When $\phi$ is admissible for $t$, a regular pseudo-term
$\phi(t)$ can be obtained by replacing each $\pa^{\prm{m}} u$ with
$\pa^{\phi^i(\prm{m})} u$, each $x^D$ with $x^{\phi(D)}$, and each
$(t)A$ with $(t)\phi(A)$.

 As for pseudo-terms there is an erasure map $(.)^-$ from p-terms
to system F terms consisting in forgetting modalities
and parameters.

A {\em linear free decoration} ({\em bang free decoration}, resp.)
of a system F type $T$ is a linear p-type (bang p-type, resp.)
$E$ such that (i) $E^- = T$,
(ii) each linear combination $\prm{c}$ occurring in $E$
consists of a single integer parameter $\prm{m}$,
and (iii) the parameters occurring in $E$ are mutually distinct.
Two free decorations $\lift{T}_1$ and $\lift{T}_2$
are said to be {\em distinct}
if the set of parameters occurring in $\lift{T}_1$ is
disjoint from the set of parameters in $\lift{T}_2$.

The {\em free decoration} $\lift{M}$ of a system F term $M$
(which is unique up to renaming of parameters) is obtained as follows:
first, to each type $T$ of a variable $x^T$ used in $M$, we associate a
bang free decoration $\lift{T}$, and to each type $U$
occurring as $(N)U$ in $T$, we associate
a linear free decoration $\lift{U}$
with the following proviso:
\begin{itemize}
\item[(i)]
one and the same $\lift{T}$ is associated
to all occurrences of the same variable $x^T$;
\item[(ii)] otherwise mutually distinct free decorations $\lift{T}_1$,
\dots, $\lift{T}_n$ are associated to
different occurrences of $T$.
\end{itemize}
$\lift{M}$ is now defined by induction on the construction of $M$:
\[
\begin{array}{rclrcl}
\lift{x^T} & = & \pa^\prm{m} x^{\lift{T}} & \\
\lift{\la x^T.M} & = & \pa^\prm{m}
\la x^{\lift{T}}. \lift{M} &
\lift{(M)N} & = & \pa^\prm{m} ((\lift{M})\lift{N}) \\
\lift{\La \alpha.M} & = & \pa^\prm{m} \La\alpha.\lift{M} &
\lift{(M)T} & = & \pa^\prm{m}((\lift{M})\lift{T})
\end{array}
\]
where all newly introduced parameters $\prm{m}$
are chosen to be fresh.
The key property of free decorations is the following:
\begin{lem}\label{instantiationlemma}
Let $M$ be a system F term and $t$ be a regular pseudo-term.
Then $t$ is a decoration
of $M$ if and only if
there is an admissible instantiation $\phi$
for $\lift{M}$ such that $\phi(\lift{M})= t$.
\end{lem}

Hence our decoration problem boils down to:
\begin{problem}[instantiation]\label{instantiationproblem}
Given a system F term $M$, determine if there exists
an admissible instantiation $\phi$ for $\lift{M}$ such that
$\phi(\lift{M})$ satisfies
the local typing, bracketing,
bang and $\Lambda$-scope conditions.
\end{problem}

 For that we
will need to be able to state the conditions of Theorem \ref{t-correct}
on p-terms;
they will yield some constraints on parameters. We will speak
of \textit{linear inequations}, meaning in fact both linear equations
and linear inequations.

\subsection{Unification constraints}
 To express the unifiability of two p-types 
$E_1$ and $E_2$, we define a set
$\unif{E_1}{E_2}$ of constraints by
\begin{eqnarray*}                                    
\unif{\alpha}{\alpha} &=& \emptyset,\\
\unif{D_1 \fm A_1}{D_2 \fm A_2}
&=& \unif{D_1}{D_2} \cup \unif{A_1}{A_2},\\
\unif{\forall \alpha.A_1}{\forall \alpha.A_2} &=&\unif{A_1}{A_2},\\
\unif{\pa^\prm{c_1} F_1}{\pa^\prm{c_2} F_2} &=&
\{\prm{c_1}=\prm{c_2}\} \cup
 \unif{F_1}{F_2},\\
\unif{\pa^{\prm{b_1},\prm{c_1}} F_1}{\pa^{\prm{b_2},\prm{c_2}} F_2} &=&
\{\prm{b_1}=\prm{b_2},\prm{c_1}=\prm{c_2}\} \cup \unif{F_1}{F_2},
\end{eqnarray*}
and undefined otherwise.
It is straightforward to observe:
\begin{lem}
 Let $E_1$, $E_2$ be two p-types such that
$\unif{E_1}{E_2}$ is defined,
and $\phi$ be an admissible
instantiation for $E_1$ and $E_2$.
Then $\phi(E_1)=\phi(E_2)$ if and only if
$\phi$ is a
solution of $\unif{E_1}{E_2}$.
\end{lem}

\subsection{Local typing constraints}

For any p-type $E$,
$\mc(E)$ denotes the set
$ \{\prm{c}\geq \prm{0} : \mbox{ $\prm{c}$ occurs in
$E$}\}$ $\cup$
$ \{\prm{b}=\prm{1}\Rightarrow\prm{c}\geq \prm{1} : \mbox{
$\pa^{\prm{b},\prm{c}} F$ occurs in $E$}\}$.
Then $\phi$ is admissible for $E$ if and only if
$\phi$ is a solution of $\mc(E)$.

When $A$ is a linear p-type $\pa^{\prm{c}} F$,
$B[A/\alpha]$ denotes a p-type obtained by
replacing each $\pa^{\prm{c'}}\alpha$ in $B$ with
$\pa^{\prm{c'}+\prm{c}}F$ and
each $\pa^{\prm{b},\prm{c'}}\alpha$ with
$\pa^{\prm{b},\prm{c'}+\prm{c}}F$.

Now consider the free decoration $\lift{M}$ of
a system F typed term $M$.
We assign
to each subterm $t$
of $\lift{M}$ a {\em linear} p-type $B$
(indicated as $t_B$)
and a set $\mc(t)$ of
constraints as on Figure \ref{localtypingconstraints}. 
Notice that any linear p-type is of the form
$\pa^{\prm{c}} F$. Moreover, since $t$ comes from 
a system F typed term, we know that
$F$ is an implication 
when $t$ occurs as $(t_{\pa^{\prm{c}} F})u$, and
$F$ is a quantification when
$t$ occurs as $(t_{\pa^{\prm{c}} F})A$.
The unification $\unif{D^\circ}{A}$ used in $\mc((t)u)$
is always defined, 
and finally,
$\lift{M}$ satisfies the eigenvariable condition.

\begin{figure*}[ht]
\[
\begin{array}{crcl}
(x^D)_{D^\circ} & \mc(x) & = & \mc(D) \\
(\pa^\prm{m} t_{\pa^\prm{c}F})_{\pa^{\prm{m}+\prm{c}}F} &
\mc(\pa^\prm{m} t) & = & \{\prm{m}+\prm{c}\geq 0\}\cup \mc(t)\\
(\la x^{D}. t_A)_{\pa^{\prm{0}} (D \fm A)} &
\mc(\la x^{D}. t) & = & \mc(D)\cup \mc(t) \\
((t_{\pa^{\prm{c}} (D\fm B)}) u_{A})_B &
\mc((t) u) & = &
\{\prm{c}=\prm{0}\}\cup \unif{D^\circ}{A}\cup
\mc(t)\cup\mc(u) \\
(\La \alpha.t_A)_{\pa^{\prm{0}}\forall\alpha. A}
& \mc(\La \alpha.t) & = & \mc(t) \\
((t_{\pa^{\prm{c}}\forall\alpha. B})A)_{B[A/\alpha]} &
\mc((t)A) & = & \{\prm{c}=\prm{0}\}\cup\mc(A)\cup \mc(t)
\end{array}
\]

 \caption{$\mc(t)$ constraints.}\label{localtypingconstraints}
\end{figure*}

Let $\Ltype(\lift{M})$ be the set
$\mc(\lift{M})\cup\{ \prm{b}=\prm{1} : $
$x^{\pa^{\prm{b},\prm{c}}F}$ occurs more than once in $\lift{M}\}$.


\subsection{Boxing constraints}
 In this section we need to recall some definitions from
 \cite{BaillotTerui05}.
We consider the words over integer parameters $\prm{m}$, $\prm{n}$
\dots, whose set we denote by $\mathcal{L}_p$.

Let $t$ be a p-term
and
$u$ an occurrence of subterm of $t$.
We define, as for pseudo-terms, the word
 $\lis{t}{u}$ in
$\mathcal{L}_p$ as follows:
$$\begin{array}{lcl}
\mbox{if $t=u$:}&& \lis{t}{u} = \epsilon, \\
\mbox{else: } &&\\
\lis{\pa^\prm{m} t}{u} &=& \prm{m}::(\lis{t}{u}) \\
 \lis{\la y^{D}.t_1}{u} &=&
\lis{\La \alpha.t_1}{u} \\
&=& \lis{(t_1)A}{u} =
\lis{t_1}{u} \\
\lis{(t_1)t_2}{u} &=& \lis{t_i}{u} \mbox{ when $t_i$ is the}\\
 && \mbox{subterm
containing $u$}.
\end{array}
$$
The sum $s(l)$ of an element $l$ of $\mathcal{L}_p$ is a linear combination
of integer parameters
defined by:
$$s(\epsilon) = \prm{0}, \quad s(\prm{m} :: l) = \prm{m}+ s(l).$$

For each  list $l\in \mathcal{L}_p$,
define $\wbracket (l) = \{ s(l')\geq \prm{0}\ |\ l'\leq l\}$
and $\bracket (l) = \wbracket (l) \cup \{s(l)=\prm{0}\}$.

Given a system F term $M$,
we define the
following sets of constraints:\\
\textbf{Bracketing constraints}.
$\Bra(\lift{M})$ is the union of the following sets:
\begin{itemize}
\item[(i)] $\bracket(\lis{\lift{M}}{x})$ for each free variable
$x$ in $\lift{M}$,
\end{itemize}
and for each occurrence of an abstraction subterm $\la x.v$ of $\lift{M}$,
\begin{itemize}
\item[(ii)] $\wbracket(\lis{\lift{M}}{\la x.v})$,
\item[(iii)] $\bracket(\lis{v}{x})$ for
each occurrence of $x$ in $v$.
\end{itemize}
\textbf{Bang constraints}.
A subterm $u_A$ that occurs as
$(t_{\pa^{\prm{c'}}(\pa^{\prm{b},\prm{c}} F\fm B)})u_A$ in $\lift{M}$
is called a {\em bang subterm} of $\lift{M}$ with the {\em critical
parameter} $\prm{b}$. Now
$\Bang(\lift{M})$ is the union of the following sets:
for each bang subterm
$u$ of $\lift{M}$ with a critical parameter $\prm{b}$,
\begin{itemize}
\item[(i)] $\{\prm{b}=\prm{0}\}$ if $u$ contains strictly more than
one occurrence of free variable, and $\{\prm{b}=\prm{1} \Rightarrow
\prm{b'}=\prm{1}\}$ if $u$ contains exactly one occurrence of free
variable $x^{\pa^{\prm{b'},\prm{c'}} F'}$.
\item[(ii)]
$\{\prm{b}=1 \Rightarrow s(\lis{u}{v})\geq \prm{1}\ :\ $
$v$ is a subterm of $u$ such that $v\neq u$ and $v\neq x\}$.
\end{itemize}
\textbf{$\La$-Scope constraints}.  
$\Scope (\lift{M})$ is the union of the following sets:
\begin{itemize}
\item $\wbracket (\lis{u}{v})$
for each subterm $\La \alpha.u$ of $\lift{M}$ and
 for each subterm $v$ of $u$ that depends on $\alpha$.
\end{itemize}

We denote
 $\Const(\lift{M})=\Ltype(\lift{M})\cup
\Bra (\lift{M})\cup \Bang (\lift{M}) \cup \Scope (\lift{M})$.
We then have:


\begin{theo}\label{t-param}
Let $M$ be a system F term and $\phi$ be an instantiation for
$\lift{M}$.
Then: $\phi$ is admissible for $M$ and
$\phi(\lift{M})$ satisfies the local typing,
bracketing, bang and $\Lambda$-scope conditions if and only if
$\phi$ is a solution of $\Const(\lift{M})$.

Moreover,
the number of (in)equations in $\Const(\lift{M})$ is quadratic
in the size of $M$.
\end{theo}

\section{Solving the constraints}

From a proof-net point of view, naively one might expect that
finding a DLAL decoration could be decomposed into first finding a
suitable EAL decoration (that is to say a box structure) and then
determining which boxes should be $\bs$ ones. This however cannot be
turned into a valid algorithm because there can be an infinite number of
EAL decorations in the first place.

 Our method will thus proceed in the opposite way: first solve the
boolean constraints, which corresponds to determine which $\bs$-boxes
are necessary, and then complete the decoration by finding a suitable
box structure.

\subsection{Solving boolean constraints}
We divide $\Const(\lift{M})$ into three disjoint sets
$\Const^b(\lift{M})$, $\Const^i(\lift{M})$ and $\Const^m(\lift{M})$:
\begin{itemize}
\item A {\em boolean constraint} $\prm{s}\in\Const^b(\lift{M})$ consists
of only boolean parameters. $\prm{s}$ is of one of the following forms:\\
\begin{tabular}{ll}
$\prm{b_1} = \prm{b_2}$ & (in $\Ltype(\lift{M})$) \\
$\prm{b} = \prm{1}$ & (in $\Ltype(\lift{M})$)\\
$\prm{b} = \prm{0}$ & (in $\Bang(\lift{M})$)\\
$\prm{b} = \prm{1}\Rightarrow
\prm{b'} = \prm{1}$ &  (in $\Bang(\lift{M})$)
\end{tabular}
\item A {\em linear constraint} $\prm{s}\in\Const^i(\lift{M})$ deals
  with integer parameters only. A linear constraint
  $\prm{s}$ is  of one of the following forms:\\
\begin{tabular}{ll}
  $\prm{c_1} = \prm{c_2}$ & (in $\Ltype(\lift{M})$) \\
  $\prm{c} \geq \prm{0}$ & (in $\Ltype(\lift{M})$, $\Bra(\lift{M})$,
  $\Scope(\lift{M})$)\\
  $\prm{c} = \prm{0}$ & (in $\Ltype(\lift{M})$ and  $\Bra(\lift{M})$)
\end{tabular}
\item A {\em mixed constraint} $\prm{s}\in\Const^m(\lift{M})$ contains
  a boolean parameter and a linear combination and is of the following
  form:\\
\begin{tabular}{ll}
  $\prm{b} = \prm{1}\Rightarrow \prm{c} \geq \prm{1}$
  & (in $\Ltype(\lift{M})$ and $\Bang(\lift{M})$)
\end{tabular}
\end{itemize}

We consider the set of instantiations on boolean parameters and the
extensional order $\leq$ on these maps: $\psi^b \leq \phi^b$ if for
any $\prm{b}$, $\psi^b(\prm{b}) \leq \phi^b(\prm{b})$.

\begin{lem}\label{minimalsolution}
  $\Const^b(\lift{M})$ has a solution if and only if it has a minimal
  solution $\psi^b$. The latter can be computed in time polynomial in
  the number of boolean constraints in $\Const^b(\lift{M})$.
\end{lem}
\begin{proof} Assuming that $\Const^b(\lift{M})$ has a solution, we can
compute the minimal one by a standard resolution procedure.
See Appendix \ref{s-proof}.\QED
\end{proof} 
%
%

\subsection{Solving integer constraints}

When $\phi^{b}$ is a boolean instantiation, 
$\phi^{b}\Const^m(\lift{M})$ denotes
the set of linear constraints defined as follows:
for any constraint of the form $\prm{b} = \prm{1}\Rightarrow
\prm{c} \geq \prm{1}$ in $\Const^m(\lift{M})$, $\prm{c} \geq
\prm{1}$ belongs to $\phi^{b}\Const^m(\lift{M})$
 if and only if $\phi^b (\prm{b})=1$.
It is then clear that 
(*) $(\phi^b, \phi^i)$ is a solution of
$\Const(\lift{M})$ if and only if $\phi^b$ is a solution of
$\Const^b(\lift{M})$ and $\phi^i$ is a solution of
$\phi^{b}\Const^m(\lift{M})\cup\Const^i(\lift{M})$.

\begin{prop}\label{solutionwithminimalbooleans}
  $\Const(\lift{M})$ admits a solution if and only if it has a
  solution $\psi=(\psi^{b}, \psi^{i})$ such that $\psi^{b}$ is the
  minimal solution of $\Const^b(\lift{M})$.
\end{prop}
\begin{proof}
Suppose that   $\Const(\lift{M})$ admits a solution $(\phi^b, \phi^i)$.
Then by the previous lemma, there is a minimal solution $\psi^b$
of $\Const^b(\lift{M})$. Since $\psi^b\leq \phi^b$, we have
  $\psi^{b}\Const^m(\lift{M})\subseteq 
\phi^{b}\Const^m(\lift{M})$. Since $\phi^i$
is a solution of $\phi^{b}\Const^m(\lift{M})\cup\Const^i(\lift{M})$
by (*) above, it is also a solution of 
$\psi^{b}\Const^m(\lift{M})\cup\Const^i(\lift{M})$.
This means that $(\psi^b, \phi^i)$ is a solution of 
$\Const(\lift{M})$.\QED
\end{proof}

Coming back to the proof-net intuition, Proposition
\ref{solutionwithminimalbooleans} means that given a syntactic tree of
term there is a most general (minimal) way to place $\bs$ boxes (and
accordingly $\bs$ subtypes in types), that is to say: if there is a
DLAL decoration for this tree then there is one with precisely this
minimal distribution of $\bs$ boxes.

Now notice that 
$\psi^{b}\Const^m(\lift{M})\cup\Const^i(\lift{M})$ is a linear
inequation system, for which a polynomial time procedure for searching
a rational solution is known.

\begin{lem}
  $\psi^{b}\Const^m(\lift{M})\cup\Const^i(\lift{M})$ has a solution in
  $\QQ$ if and only if it has a solution in $\ZZ$.
\end{lem}

\begin{proof}
  Clearly the set of solutions is closed under multiplication by a
  positive integer.\QED
\end{proof}

\begin{theo}\label{t-const}
  Let $M$ be a System F term. Then one can decide in time polynomial
  in the number of constraints in $\Const(\lift{M})$ whether
  $\Const(\lift{M})$ admits a solution.
\end{theo}

\begin{proof}
  First apply the procedure described in the proof of Lemma
  \ref{minimalsolution} to decide if there is a minimal solution
  $\psi^b$ of $\Const^b(\lift{M})$. If it exists, apply the polynomial
  time procedure to decide if
  $\psi^b\Const^m(\lift{M})\cup\Const^i(\lift{M})$ admits a solution
  in $\QQ$. If it does, then we also have an integer solution.
  Otherwise, $\Const(\lift{M})$ is not solvable.\QED
\end{proof}

By combining Theorem \ref{t-correct}, Lemma \ref{instantiationlemma},
Theorems \ref{t-param} and \ref{t-const}, we obtain our main theorem:
\begin{theo}
Given a system F term $M^T$, it is decidable in
time polynomial in the size of $M$ whether
there is a decoration $A$ of $T$ such that
$\vdash_{DLAL} M: A$.
\end{theo}

\section{Implementation}\label{l-implement}

\subsection{Overview}

We designed an implementation of the type inference algorithm. The
program is written in functional Caml and is quite concise (less than
1500 lines). A running program not only shows the actual feasibility
of our method, but also is a great facility for building examples, and
thus might allow for a finer study of the algorithm.

Data types as well as functions closely follow the previous
description of the algorithm: writing the program in such a way tends
to minimise the number of bugs, and speaks up for the robustness of
the whole proof development.

The program consists of several successive parts:
\begin{enumerate}
\item{Parsing phase: turns the input text into a concrete syntax
    tree. The input is an F typing judgement, in a syntax \emph{\`a la}
    Church with type annotations at the binders. It is changed into the de
    Bruijn notation, and parameterized with fresh
    parameters. Finally, the abstract tree is decorated with
    parameterized types at each node.}
\item{Constraints generation: performs explorations on the tree
    and generates the boolean, linear and mixed constraints.}
\item{Boolean constraints resolution: gives the minimal solution of the
boolean constraints, or answers negatively if
    the set admits no solution.}
\item{Constraints printing: builds the final set of linear constraints.}
\end{enumerate}

We use the simplex algorithm to solve the linear constraints. It runs in
$O(2^n)$, which comes in contrast with the previous result of
polynomial time solving, but has proven to be the best in practice
(with a careful choice of the objective function).


\subsection{An example of execution}
As an example, let us consider the reversing function \textbf{rev} on binary words,
applied to \textbf{1010}. \textbf{rev} can be defined by a single higher-order
iteration, and thus represented by the following system F term:
$$
\begin{array}{l}
  \lambda l^{W}.\Lambda \beta.\lambda so^{\beta \rightarrow \beta}.\lambda si^{\beta \rightarrow \beta}. (l \; (\beta \rightarrow \beta))\\ 
   \; \; \; \; \lambda a^{\beta \rightarrow \beta}.\lambda x^{\beta}.(a) (so) x\\
   \; \; \; \; \lambda a^{\beta \rightarrow \beta}.
   \lambda x^{\beta}.(a) (si) x \; (\Lambda \alpha.\lambda z^{\alpha}.z) \beta
 \end{array}
 $$
We apply it to :
$$ \Lambda \alpha.\lambda so^{\alpha \rightarrow \alpha}.\lambda si^{\alpha \rightarrow \alpha}.\lambda x^{\alpha}.(si) (so) (si) (so) x, $$
representing the word \textbf{1010}.
Since \textbf{rev} involves higher-order functionals and polymorphism, 
it is not so straightforward to tell,
just by looking at the term structure, 
whether it works in polynomial time or not.

Given \textbf{rev(1010)} as input (coded by ASCII characters),
our program produces 
177 (in)equations on 79 variables. After constraint solving,
we obtain the result, that can be read as:
$$
\begin{array}{l}
 (\lambda l^{W}.
   \Lambda \beta . \lambda so^{\bs (\beta \llto \beta)} . \lambda si^{\bs (\beta \llto \beta)}.\\
   \; \; \; \; \; \; \; \; \pa (\pad ((l \; (\beta \llto \beta)) \\
   \; \; \; \; \; \; \; \; \pa \lambda a^{\beta \llto \beta}.\lambda x^{\beta}.(a) (\pad so) x \\
   \; \; \; \; \; \; \; \; \pa \lambda a^{\beta \llto \beta}.\lambda x^{\beta}.(a) (\pad si) x)\\
   \; \; \; \; \; \; \; \; (\Lambda \alpha.\lambda z^{\alpha}.z) \beta) \;\\
   \Lambda \alpha.\lambda so^{\bs\alpha \rightarrow \alpha}.\lambda si^{\alpha \rightarrow \alpha}.\pa \lambda x^{\alpha}.(\pad si) (\pad so) (\pad si) (\pad so) x
 \end{array}
$$
 It corresponds to the natural depth-1 typing of this term, with
conclusion type $W_{DLAL}\fm W_{DLAL}$. 
 The solution ensures polynomial time termination, and 
in fact its depth guarantees normalization in a quadratic
number of $\beta$-reduction steps.

Further examples, as well as the program itself, will be available 
at
\begin{center}
 \texttt{http://www-lipn.univ-paris13.fr/\~{ }atassi/}
\end{center}
 \bibliographystyle{alpha}
  \bibliography{infdlal}

\newpage

\appendix

\begin{center} APPENDIX\end{center}
\section{Proofs}\label{s-proof}

\noindent
{\em Proof of Proposition \ref{p-correct}.}

\begin{proof}
First, one can build a (possibly non-regular)
decoration $M^+$ of $M$ satisfying the four conditions
by induction on the derivation.
Depending on the last typing rule used (see Figure \ref{NDLALrules}), 
$M^+$ takes one of the following forms:
$$
\begin{array}{llll}
\mbox{(Id)} & x^{A^\star} & 
\mbox{($\pa$ e)} & M^+[N^+/x] \\
\mbox{($\fm$ i)} & \la x^{A^\star}.M^+  &
\mbox{($\fm$ e)} & (M^+)N^+ \\
\mbox{($\fli$ i)} & \la x^{\bs A^\star}.M^+  &
\mbox{($\fli$ e)} & (M^+)\pa N^+[\pad z^{\bs C^\star}/z] \\
\mbox{(Weak)} & M^+  &
\mbox{(Cntr)} & M^+[x/x_1, x/x_2] \\
\mbox{($\forall$ i)} & \La \alpha.M^+  &
\mbox{($\forall$ e)} & (M^+)B^\star \\
\mbox{($\pa$ i)} & 
\multicolumn{3}{l}{
\pa M^+ [\pad x_i^{\bs A_i^\star}/ x_i, \pad y_j^{\pa B_j^\star}/ y_j]}
\end{array}
$$
where $M^+$ in ($\pa$ i) has free variables
$x_1^{A_1}, \dots,$ $x_m^{A_m},$ $y_1^{B_1}, \dots,$ $y_n^{B_n}$.

It is easily verified that $M^+$ has a suitable type and satisfies the 
four conditions; let us just remark:
\begin{itemize}
\item The bang condition for $(M^+)\pa N^+[\pad z^{\bs C^\star}/z]$
in ($\fli$ e)
follows by the bracketing condition for $N^+$, which holds by the induction
hypothesis, while the $\La$-scope condition follows by
the eigenvariable condition for $N^+$.
Similarly for the case of ($\pa$ i).
\item $M^+[N^+/x]$ in ($\pa$ e)
satisfies the $\La$-scope condition since
substitution is capture-free, and satisfies the bang condition
since $x$ has a linear type and thus cannot appear inside a bang 
subterm of $M^+$.
\end{itemize}

Finally, the required  regular pseudo-term $t$ is obtained from 
$M^+$ by applying inside $t$ the following rewrite rules
as many times as possible:
$$
\pad\pa u\ \longrightarrow\ u, \quad\quad
\pa\pad u\ \longrightarrow u.$$
It is clear that the four conditions are preserved by these
reductions.\QED
\end{proof}
\medskip

\noindent
{\em Proof of Lemma \ref{boxinglemma}.}

\begin{proof}
Given $\pa t$, assign an index to
each occurrence of $\pa$ and $\pad$ in $\pa t$
to distinguish occurrences
(we assume that
the outermost $\pa$ have index $0$).
One can then
find closing brackets $\pad_1, \dots, \pad_{n}$
that match the opening
bracket $\pa_0$ in $\pa_0 t$.
Replace each $(\pad_i u_i)_{B_i}$ with
a fresh and distinct free variable $x_i^{B_i}$ ($1\leq j\leq n$),
and let $\pa v$ be the resulting pseudo-term.
This way one can obtain
$v$, $u_1$, \dots, $u_n$,
 such that condition 2 holds.

As to point 1., we claim that $v$ does not contain a free variable
other than $x_1, \dots, x_n$.  If there is any, say $y$, then it is
also a free variable of $t$, thus the bracketing condition for $\pa_0
t$ implies that $\lis{\pa_0 t}{y}$ is well-bracketed, and thus there
is a closing bracket that matches $\pa_0$ in the path from $\pa_0 t$
to $y$.  That means that $y$ belongs to one of $u_1$, \dots, $u_n$,
not to $v$.  Hence condition 1 holds.

We now need to check point 3. The bracketing condition for $v$, $u_1, \dots, u_n$
can be shown as in
\cite{BaillotTerui05}.
The $\La$-scope condition is easy to verify.

As to the local typing condition, the only nontrivial point is
that $v$ satisfies the eigenvariable condition.
Suppose that the type $B_i$ of $x_i$ contains a bound variable $\al$
of $v$. Then $\pa_0 t$ contains a subterm
of the form $\La \al.v'[\pad_i u_i/x_i]$ and $u_i$ depends on $\alpha$.
However, $\lis{v''}{u_i}$
with $v'' = v'[\pad_i u_i/x_i]$ cannot be weakly well-bracketed
because $\pad_i$ should match the outermost opening bracket
$\pa_0$. This contradicts the $\La$-scope condition
for $\pa_0 t$.

To show the bang condition for $v$ (it is clear for $u_1, \dots,
u_n$), suppose that $v$ contains a bang subterm $v'$.  We claim that
$v'$ does not contain variables $x_1, \dots, x_n$.  If it contains
any, say $x_i$, then $\pa_0 t$ contains $v'[\pad_i u_i/x_i]$ and the
bang condition for $\pa_0 t$ implies that $s(\lis{v''}{\pad u_i})\geq
1$ with $v''=v'[\pad_i u_i/x_i]$. On the other hand, we clearly have
$s(\lis{\pa_0 t}{v''})\geq 1$ because $v''$ contains the closing
bracket $\pad_i$ that matches $\pa_0$.  As a consequence, we have
$s(\lis{\pa_0 t}{\pad_i u_i})\geq 2$.  This means that $\pad_i$ does not
match $\pa_0$, a contradiction.  As a consequence, $v'$ does not
contain $x_1, \dots, x_n$. So $v'$ occurs in $\pa_0 t$, and
therefore satisfies the bang condition.\QED
\end{proof}

\medskip
{\em Proof of Theorem \ref{t-correct}.}
The `only if' direction has already been given by Proposition
\ref{p-correct}.
The other direction is proved
by induction on the size of pseudo-term $t$.

When $t$ is a variable $(x^D)_{D^\circ}$, the claim can be
established by (Id) and ($\pa$ i).
Note that $t$ cannot be of the form
$\pad u$ due to the bracketing condition.

When $t$ is one of
$\la x^{D}. u$,
$(u)v$ (with $v$ not a bang subterm),
$\La \alpha.u$, $(u)A$,
the subterms $u$ and $v$ also
satisfy all conditions. Hence we can use the induction
hypothesis to show that $t^-$ is typable in \DLAL.
When $t$ is $\pa u$, apply Lemma \ref{boxinglemma} and
argue similarly by using rules ($\pa$ e).

When $t$ is $(u_{\bs A\fm B})v_{\pa A}$, i.e., with $v$ a bang subterm,
we have $\Gamma;\Delta\vdash u^- :
A\fli B$ with suitable $\Gamma$ and $\Delta$ by the induction
hypothesis.

If $v$ is a variable, then it must be of the form
$x^{\bs A}$ by the bang condition (i).
Hence by applying ($\fli$ e)
to
$\Gamma;\Delta\vdash u^- : A\fli B$ and
$;x:A\vdash x:A$, we obtain
$\Gamma, x:A; \Delta\vdash (u^-)x : B$ as required.

If $v$ is not a variable, then it must be
of the form $\pa v_0$ due to the bang condition (ii)
and contain at most one free variable. Let us suppose
that it contains $y^{\bs C}$.
Now, the bracketing condition implies
$s(\lis{\pa v_0}{y})=0$ while the bang condition implies
$s(\lis{\pa v_0}{v'})\geq 1$
for any subterm $v'$ of $v_0$ other than $y$.
Therefore, combined with Lemma \ref{boxinglemma}, it follows that
$v$ is actually of the form $\pa v_1 [\pad y/x]$,
where $v_1$ contains a variable $x^C$ and satisfies all the
conditions. By induction hypothesis, we have
$; x:C\vdash v_1^- : A$, and hence $; y:C\vdash v_1^-[y/x] : A$
by renaming.
Therefore, we obtain
$\Gamma, y:C;\Delta\vdash (u^-) v^- : B$ by ($\fli$ e).\QED\\

 
\noindent
{\em Proof of Lemma \ref{minimalsolution}}.
 Let $\bc :=\Const^b(\lift{M})$.  Apply repeatedly the following
  steps until reaching a fixpoint:
\begin{itemize}
\item if $\prm{b_1}=\prm{b_2} \in \bc $ and $\prm{b_1}=\prm{0} \in \bc $
(resp.  $\prm{b_1}=\prm{1} \in \bc $), then let
$\bc := \bc  \cup \{\prm{b_2}=\prm{0}\}$
(resp. $\bc := \bc  \cup \{\prm{b_2}=\prm{1}\}$);
\item if $ (\prm{b}=\prm{1} \Rightarrow \prm{b'}=\prm{1})
\in \bc $ and $\prm{b}=\prm{1}
\in \bc $, then let $\bc := \bc  \cup \{\prm{b'}=\prm{1}\}$.
\end{itemize}
 It is obvious that this can be done in a polynomial number of steps and
that the resulting system $\bc $ is equivalent to $\Const^b(\lift{M})$.

Now, if $\bc $ contains a pair of equations: $\prm{b}=0, \prm{b}=1$, then
it is inconsistent. Otherwise define the boolean instantiation
$\psi^{b}$ such that $\psi^{b}(\prm{b}) := 1$ if $\prm{b}=1 \in \bc $ and $\psi^{b}(\prm{b}):=0$ otherwise:

It is clear that $\psi^{b}$ is a solution of $\bc $. In particular,
observe that any constraint of the form $(\prm{b}=1 \Rightarrow
\prm{b'}=1)$ in $\bc $ is satisfied by $\psi^{b}$. Moreover any
solution $\phi^{b}$ of $\bc $ satisfies: $\psi^b \leq \phi^b$.
Therefore if $\Const^b(\lift{M})$ has a solution then it has a minimal
one.\QED

\end{document}